\documentclass[12pt]{article}
\def\be{\begin{equation}}
\def\ee{\end{equation}}
\def\bea{\begin{eqnarray}}
\def\eea{\end{eqnarray}}
\usepackage{graphicx}

\catcode`\@=11
\def\lsim{\mathrel{\mathpalette\@versim<}}
\def\gsim{\mathrel{\mathpalette\@versim>}}
\def\@versim#1#2{\vcenter{\offinterlineskip
\ialign{$\m@th#1\hfil##\hfil$\crcr#2\crcr\sim\crcr } }}
\catcode`\@=12
\usepackage{axodraw}

\catcode`@=12
\begin{document}
\thispagestyle{empty}
\begin{flushright}
UCRHEP-T456\\
October 2008\
\end{flushright}
\vspace{0.3in}
\begin{center}
{\LARGE \bf Seesaw Neutrino Mass and\\
New U(1) Gauge Symmetry\\}
\vspace{0.8in}
{\bf Rathin Adhikari$^a$, Jens Erler$^b$, and Ernest Ma$^c$\\}
\vspace{0.2in}
{\sl $^a$ Centre for Theoretical Physics, Jamia Millia Islamia -- Central 
University,\\ Jamia Nagar, 
New Delhi 110025, India\\} 
\vspace{0.1in}
{\sl $^b$ Instituto de Fisica, Univeridad Nacional Autonoma de Mexico,\\ 
01000 Mexico, D. F., Mexico\\} 
\vspace{0.1in}
{\sl $^c$ Department of Physics and Astronomy, University of California,\\ 
Riverside, California 92521, USA\\}
\end{center}
\vspace{0.8in}
\begin{abstract}\
The three electroweak doublet neutrinos $\nu_{e,\mu,\tau}$ of the Standard Model 
may acquire small seesaw masses, using either three Majorana fermion singlets 
$N$ or three Majorana fermion triplets $(\Sigma^+,\Sigma^0,\Sigma^-)$. It is 
well-known that the former accommodates the U(1) gauge symmetry $B-L$.  
It has also been shown some years ago that the latter supports a new 
$U(1)_X$ gauge symmetry.  Here we study two variations of this $U(1)_X$, 
one for two $N$ and one $\Sigma$, the other for one $N$ and two $\Sigma$.  
Phenomenological consequences are discussed.
\end{abstract}
   
\newpage
\baselineskip 24pt

\noindent \underline{\it Introduction} : With the observation of neutrino 
oscillations, the question of neutrino mass is at the forefront of many 
theoretical studies in particle physics.  A minimal (and essentially trivial) 
solution is to add three neutral fermion 
singlets $N_R$ (commonly referred to as right-handed neutrinos) so that the 
famous canonical seesaw mechanism, i.e. $m_\nu \simeq -m_D^2/m_N$, is 
realized, where $m_D$ is the Dirac mass linking $\nu_L$ to $N_R$ and $m_N$ 
is the heavy Majorana mass of $N_R$.  On the other hand, this is not the 
only way to realize the generic seesaw mechanism which is implicit in the 
unique dimension-five effective operator \cite{w79}
\begin{equation}
{\cal L}_5 = - {f_{ij} \over 2 \Lambda} (\nu_i \phi^0 - l_i \phi^+) (\nu_j 
\phi^0 - l_j \phi^+) + H.c.
\end{equation}
for obtaining small Majorana masses in the standard model (SM) of particle 
interactions.  In fact, there are three tree-level (and three generic 
one-loop) realizations \cite{m98}.  The second most often considered 
mechanism for neutrino mass is that of a scalar triplet 
$(\xi^{++},\xi^+,\xi^0)$, whereas the third tree-level realization, i.e. that 
of a fermion triplet $(\Sigma^+,\Sigma^0,\Sigma^-)$ \cite{flhj89}, has not 
received as much attention.  However, it may be essential for gauge-coupling 
unification \cite{m05,bs07,df07,ms08} in the SM, and be probed 
\cite{bns07,fhs08,aa08} at the Large Hadron Collider (LHC).  
It is also being discussed in a variety of other contexts 
\cite{f07,abbgh08,ff08,m08-4,moy08}.  A new U(1) gauge symmetry 
\cite{m02,mr02,bd05} is another remarkable possibility, and in this paper 
we study in some detail two versions of this extension, one with two $N$ 
and one $\Sigma$, the other one $N$ and two $\Sigma$.

\begin{table}[htb]
\caption{Fermion content of proposed model.}
\begin{center}
\begin{tabular}{|c|c|c|}
\hline 
Fermion & $SU(3)_C \times SU(2)_L \times U(1)_Y$ & $U(1)_X$ \\ 
\hline
$(u,d)_L$ & $(3,2,1/6)$ & $n_1$ \\ 
$u_R$ & $(3,1,2/3)$ & $n_2$ \\
$d_R$ & $(3,1,-1/3)$ & $n_3$ \\
$(\nu,e)_L$ & $(1,2,-1/2)$ & $n_4$ \\
$e_R$ & $(1,1,-1)$ & $n_5$ \\
\hline
$N_R$ & $(1,1,0)$ & $n_6$ \\
$(\Sigma^+,\Sigma^0,\Sigma^-)_R$ & $(1,3,0)$ & $n_6$ \\
\hline
\end{tabular}
\end{center}
\end{table}

\noindent \underline{\it New U(1) gauge symmetry} :
Consider the fermions of the SM plus $N$ and $\Sigma$ under a new $U(1)_X$ 
gauge symmetry as listed in Table 1.  To obtain masses for all the quarks 
and leptons, four Higgs doublets $\Phi_i = (\phi^+,\phi^0)_i$ with $U(1)_X$ 
charges $n_1-n_3$, $n_2-n_1$, $n_4-n_5$, and $n_6-n_4$ are required, but some 
of these may turn out to be the same, depending on the anomaly-free solutions 
of $n_i$ to be discussed below.  To obtain large Majorana masses for $N$ and 
$\Sigma$, and to break $U(1)_X$ spontaneously, the Higgs singlet $\chi^0$ 
with $U(1)_X$ charge $-2n_6$ or $2n_6$ will also be required.

Assuming three families of quarks and leptons and the number of $N$ and 
$\Sigma$ to be $n_N$ and $n_\Sigma$ with $n_N + n_\Sigma = 3$, we consider the 
conditions for the absence of the axial-vector anomaly \cite{a69,bj69,b69} 
in the presence of $U(1)_X$ \cite{m02}.
\bea
[SU(3)]^2 U(1)_X &:& 2n_1 - n_2 - n_3 = 0, \\ 
{[SU(2)]}^2 U(1)_X &:& (9/2) n_1 + (3/2) n_4  - 2 n_\Sigma n_6 = 0, \\ 
{[U(1)_Y]}^2 U(1)_X &:& (1/6) n_1 - (4/3) n_2 - (1/3) n_3 + (1/2) n_4 -  
n_5 = 0, \\ 
U(1)_Y [U(1)_X]^2 &:& n_1^2 - 2 n_2^2 + n_3^2 - n_4^2 + n_5^2 = 0, \\ 
{[U(1)_X]}^3 &:& 3 [6 n_1^3 - 3 n_2^3 - 3 n_3^3 + 2 n_4^3 - n_5^3] - 
(3 n_\Sigma + n_N) n_6^3 = 0.
\end{eqnarray}
Furthermore, the absence of the mixed gravitational-gauge anomaly 
\cite{ds72,ef76,aw84} requires the sum of $U(1)_X$ charges to vanish, i.e.
\be
U(1)_X : 3 [6 n_1 - 3 n_2 - 3 n_3 + 2 n_4 - n_5] - (3 n_\Sigma + n_N) n_6 = 0.
\ee
Since the number of $SU(2)_L$ doublets remains even (it is in fact unchanged), 
the global SU(2) chiral gauge anomaly \cite{w82} is absent automatically.

Equations (2), (4), and (5) do not involve $n_6$.  Together they allow two 
solutions:
\be
{\rm(I)}~~n_4=-3n_1, ~~~~ {\rm(II)}~~n_2 = (7n_1-3n_4)/4.
\ee
In the case of solution (I), if $n_\Sigma \neq 0$, then Eq.~(3) implies 
$n_6=0$, from which it can easily be seen that $U(1)_X$ is proportional 
to $U(1)_Y$, i.e. no new gauge symmetry is obtained.  If $n_\Sigma = 0$, 
then $n_3 = 2n_1-n_2$ and $n_5 = -2n_1-n_2$, and Eqs.~(6) and (7) become
\bea
3(-4n_1+n_2)^3-n_N n_6^3 &=& 0, \\
3(-4n_1+n_2)-n_N n_6 &=& 0.
\eea
For $n_N = 3$, we obtain $n_6 = -4n_1+n_2$ which has two independent 
solutions:  $n_1 = 1/6$ and $n_2 = 2/3$ imply $U(1)_Y$, whereas $n_1 = n_2 
= 1/3$ imply $U(1)_{B-L}$ as is well-known.  In the case of solution (II),
\be
n_3 = (n_1+3n_4)/4, ~~~~ n_5 = (-9n_1+5n_4)/4,
\ee
and Eq.~(3) yields
\be
n_6 = {3 \over 4 n_\Sigma} (3n_1+n_4).
\ee
Equations (6) and (7) become
\bea
&& 9(3n_1+n_4)^3/64 - (3n_\Sigma + n_N) n_6^3 = 0, \\
&& 9(3n_1+n_4)/4 - (3n_\Sigma + n_N) n_6 = 0.
\eea
The unique solution is thus $n_N = 0$ and $n_\Sigma = 3$.  However, if we 
insist that $n_N = 3 - n_\Sigma \neq 0$, then the nonzero $[U(1)_X]^3$ 
and $U(1)_X$ anomalies given by $(n_\Sigma^3/3 - 2 n_\Sigma - 3) n_6^3$ 
and $(n_\Sigma - 3) n_6$ may be canceled by the addition of more singlets 
without affecting the other conditions.  For $n_\Sigma = 2$ ($n_N = 1$), 
they are $(-13/6)n_6^3$ and $-n_6$, which cannot be canceled by just one 
chiral fermion. However, a unique solution exists for two right-handed 
singlets of $U(1)_X$ charges $(-5/3)n_6$ and $(2/3)n_6$. Similarly, for 
$n_\Sigma = 1$ ($n_N = 2$), they are canceled by right-handed singlets of 
$U(1)_X$ charges $(-5/3)n_6$ and $(-1/3)n_6$.  We list in Table 2 the 
resulting four models with $n_\Sigma + n_N = 3$, where the last three columns 
correspond to the $U(1)_X$ charges of possible Higgs doublets $\Phi_{1,2,3}$ 
which couple to the quarks, charged leptons, and neutrinos, respectively.  
Note that these extra singlets $S_{1R,2R}$ are distinguished from $N_R$ 
by their $U(1)_X$ charges.  Whereas $N_R$ (and $\Sigma_R$) are chosen to be 
the seesaw anchors for the Majorana neutrino masses through their couplings 
to the lepton doublets and a Higgs doublet with the appropriate $U(1)_X$ 
charge, $S_{1R,2R}$ are not.  However, in the case of Model (C), $S_{1R}$ 
just happens to have the required $U(1)_X$ charge which lets it couple 
to the lepton doublets through the Higgs doublet which gives rise to quark 
masses.  Note also that we do not consider the exceptional case where one 
neutrino is massless, hence the number of $N_R$ plus $\Sigma_R$ is always 
set equal to three.

\begin{table}[htb]
\caption{$U(1)_X$ properties of Models (A) to (D).}
\begin{center}
\begin{tabular}{|c|c|c|c|c|c|c|}
\hline 
Model & $N_R$ & $\Sigma_R$ & $n_6$ & $n_1-n_3=n_2-n_1$ & $n_4-n_5$ & $n_6-n_4$ 
\\ 
\hline
(A) & 3 & 0 & $-4n_1+n_2$ & $n_2-n_1$ & $n_2-n_1$ & $n_2-n_1$ \\ 
\hline
(B) & 2 & 1 & $(3/4)(3n_1+n_4)$ & $(3/4)(n_1-n_4)$ & $(1/4)(9n_1-n_4)$ & 
$(1/4)(9n_1-n_4)$  \\ 
\hline
(C) & 1 & 2 & $(3/8)(3n_1+n_4)$ & $(3/4)(n_1-n_4)$ & $(1/4)(9n_1-n_4)$ & 
$(1/8)(9n_1-5n_4)$ \\ 
\hline
(D) & 0 & 3 & $(1/4)(3n_1+n_4)$ & $(3/4)(n_1-n_4)$ & $(1/4)(9n_1-n_4)$ & 
$(3/4)(n_1-n_4)$  \\ 
\hline
\end{tabular}
\end{center}
\end{table}

\noindent {\bf (A)} This is the canonical seesaw model with three singlets. 
Since the last three columns, corresponding to the $U(1)_X$ assignments 
of the Higgs doublets $\Phi_i$ required for quark, charged-lepton, and 
neutrino masses respectively, are the same, only the one standard Higgs 
doublet is required.

\noindent {\bf (D)} This is the seesaw model where $N_R$ is replaced by 
$(\Sigma^+,\Sigma^0,\Sigma^-)_R$ per family.  Two different Higgs doublets 
($\Phi_1=\Phi_3$, and $\Phi_2$) are required.

\noindent {\bf (B)} Here two $N_R$ and one $\Sigma_R$ with the same $U(1)_X$ 
assignment are present.  One Higgs doublet ($\Phi_1$) couples to quarks, the 
other ($\Phi_2=\Phi_3$) to leptons. 

\noindent {\bf (C)} Here one $N_R$ and two $\Sigma_R$ are present. Three 
different Higgs doublets are required, opening up the possibility that 
neutrino masses are radiative, in the manner proposed first in 
Ref.~\cite{m06}.

\begin{table}[htb]
\caption{$U(1)_X$ content of new particles in Model (B).}
\begin{center}
\begin{tabular}{|c|c|}
\hline 
Particle & $U(1)_X$ \\
\hline
$N_{1R}, N_{2R}, (\Sigma^+,\Sigma^0,\Sigma^-)_R$ & $(3/4)(3n_1+n_4)$ \\ 
$S_{1R}$ & $-(1/4)(3n_1+n_4)$ \\ 
$S_{2R}$ & $-(5/4)(3n_1+n_4)$ \\ 
\hline
$(\phi^+,\phi^0)_1$ & $(3/4)(n_1-n_4)$ \\ 
$(\phi^+,\phi^0)_2$ & $(1/4)(9n_1-n_4)$ \\ 
\hline
$\chi_1$ & $-(1/2)(3n_1+n_4)$ \\ 
$\chi_2$ & $-(3/2)(3n_1+n_4)$ \\ 
\hline
\end{tabular}
\end{center}
\end{table}

\noindent \underline{\it Model with one triplet} :
Consider now Model (B) in more detail.  In addition to the SM fermions, the 
other fermions and scalars are listed in Table 3.  Quarks acquire masses 
through $\Phi_1$ and leptons through $\Phi_2$.  In addition, the Yukawa terms 
$N_{R} N_{R} \chi_2$, $\Sigma_R \Sigma_R \chi_2$, $S_{1R} S_{1R} \chi_1^\dagger$, 
$S_{1R} S_{2R} \chi_2^\dagger$, $N_{R} S_{1R} \chi_1$, $N_{R} S_{2R} 
\chi_1^\dagger$ are allowed.  As $U(1)_X$ is broken spontaneously by the 
vacuum expectation values $\langle \chi_{1,2} \rangle$, all the new fermions 
acquire large Majorana masses.  As for the Higgs potential consisting of 
$\Phi_{1,2}$ and $\chi_{1,2}$, it has many allowed terms.  Two are of 
particular importance, namely $\chi_1 \Phi_1^\dagger \Phi_2$ and $\chi_1^3 
\chi_2^\dagger$, without which there would be two unwanted global U(1) 
symmetries.

The $X$ gauge boson mixes with the $Z$ boson of the SM because $\phi^0_{1,2}$ 
transform under both $SU(2)_L \times U(1)_Y$ and $U(1)_X$.  It also 
contributes directly to quark and lepton neutral-current interactions.  
Therefore, its mass and coupling are constrained by present experimental data.
This is common to Models (B), (C), and (D). 
Let $\langle \phi_{1,2}^0 \rangle = v_{1,2}$ and $\langle \chi_{1,2}^0 \rangle 
= u_{1,2}$, then the $2 \times 2$ mass-squared matrix spanning $Z$ and $X$ 
is given by
\begin{eqnarray}
&& M_{ZZ}^2 = {1 \over 2} g_Z^2 (v_1^2 + v_2^2), \\ 
&& M_{ZX}^2 = M_{XZ}^2 = {3 \over 8} g_Z g_X (n_1-n_4) v_1^2 + 
{1 \over 8} g_Z g_X (9n_1-n_4) v_2^2, \\ 
&& M_{XX}^2 = {1 \over 2} g_X^2 (3n_1 + n_4)^2 (u_1^2 + 9  u_2^2) + 
{9 \over 8} g_X^2 (n_1-n_4)^2 v_1^2 + {1 \over 8} g_X^2 (9n_1-n_4)^2 v_2^2.
\end{eqnarray}
In general, there is $Z-X$ mixing in their mass matrix, but it must be very 
small to satisfy present precision electroweak measurements.  Of course, 
increasing $M_X$ to 10 TeV or so is a possible solution, but there is also 
a condition for zero $Z-X$ mass mixing: $v_2^2/v_1^2 = 3(n_4-n_1)/(9n_1-n_4)$, 
which requires $1 < n_4/n_1 < 9$.  For example, if $v_1^2=v_2^2= v^2/2$, then  
$n_4 = 3n_1$. In that case,
\begin{equation}
M_Z^2 = (1/2) g_Z^2 v^2, ~~~ M_X^2 = 18 n_1^2 g_X^2 (u_1^2 + 9 u_2^2) 
+ (9/2) g_X^2 v^2.
\end{equation}
However, there may also be kinetic mixing \cite{h86}, unless $U(1)_Y$ and 
$U(1)_X$ are orthogonal \cite{lt99}, which is achieved with $n_4/n_1 = 13/9$. 
In that case, it may be avoided up to one loop.  For zero mass mixing, this 
then requires $v_2^2/v_1^2=3/17$.

\noindent \underline{\it Low-engergy constraints} :
Precision data at the $Z$ pole are insensitive to additional direct 
contributions to fermion pair production from the virtual $X$ boson. 
However, $Z$ pole data can be affected indirectly through $Z-X$ mixing, 
generally leading to a shift in the measured $Z$ mass and a modification of 
its couplings to SM fermions.  The high precision of these data and their 
good agreement with the SM predictions typically constrain the $Z-X$ mixing 
to be well below one percent~\cite{Erler99}.  For simplicity, we restrict 
ourselves here to the case with no mixing.

In contrast, precision measurements at energies or momentum transfers much 
below the electroweak scale can give strong constraints on the interactions 
of the $X$ boson, comparable with or stronger than collider limits from the 
Tevatron (dilepton invariant mass distribution~\cite{dilepton} and its 
forward-backward asymmetry~\cite{asymmetry}) and LEP 2~\cite{LEP2} 
(fermion pair production). At low energies, these are interference effects 
with photon exchange amplitudes which are parametrically suppressed by only 
two powers of the heavy boson mass, being proportional to $M_Z^2/M_X^2$.

In particular, the weak charge $Q_W$ of heavy nuclei as measured in 
atomic parity violation (APV) is very sensitive to extra U(1) gauge bosons.  
Most accurately known is the weak charge of cesium, where the uncertainties 
of both the APV measurements~\cite{Boulder,Paris} and the necessary many-body 
atomic structure calculations~\cite{Flambaum} are below the 0.5\% level.  
We also include $Q_W(Tl)$~\cite{Oxford,Seattle} in our analysis. 
Furthermore, there is the weak charge of the electron which has been
extracted by the E-158 Collaboration~\cite{E158} from polarized M{\o}ller 
scattering at the SLC.  For example, at the SM tree level one has 
$Q_W^e = 1 - 4 \sin^2\theta_W$, where $\theta_W$ is the weak mixing
angle.  This is modified in the presence of the $X$ boson (and in the
absence of $Z-X$ mixing), {\em viz.}, $$Q_W^e =  1 - 4 \sin^2\theta_W 
  - {g_X^2\sin^2\theta_W\cos^2\theta_W M_Z^2\over\pi\alpha M_X^2} (e_L^2
- e_R^2),$$ where $e_L = n_4$ and $e_R = (5 n_4 - 9 n_1)/4$. The weak 
charges of up and down quarks coherently building up the weak charges 
of heavy nuclei are modified in a similar way.

There are various measurements of neutrino and anti-neutrino deep
inelastic scattering (DIS) cross sections, dominated by the result of the 
NuTeV Collaboration~\cite{NUTEV}. The original NuTeV analysis~\cite{NUTEV} 
assumed a symmetric strange quark sea for the parton distribution
functions.  Subsequently, NuTeV determined the strange-quark asymmetry
experimentally and found $S^- \equiv \int_0^1 dx x [s(x)- \bar{s}(x)] =
0.00196 \pm 0.00135 \neq 0$~\cite{SSbar}. As a consequence, we used 
Ref.~\cite{NuTeV03} to adjust their value for the left-handed effective 
coupling, $g_L^2 = 0.30005 \pm 0.00137$ to $g_L^2 = 0.3010 \pm 0.0015$, 
reducing the initial deviation from the SM of almost 3~standard deviations 
by about $1\sigma$. The right-handed coupling $g_R^2$ and the older 
$\nu$-DIS results from CDHS~\cite{CDHS} and CHARM~\cite{CHARM} at CERN and 
CCFR~\cite{CCFR} at FNAL are expected to exhibit shifts due to $S^- \neq 
0$ as well, but these ought to be less significant since their relative 
experimental uncertainties are larger. For more details, see 
Ref.~\cite{PDG2008}.

At the one-loop level, the $X$ boson also contributes to anomalous magnetic
moments, but the effect is negligible relative to the experimental
uncertainties.  Finally, box diagrams containing $X$ bosons affect tests of
CKM unitarity relations, the most precise of which being $|V_{ud}|^2 + 
|V_{us}|^2 + |V_{ub}|^2  = 0.9999 \pm 0.0006$~\cite{Marciano}.  These effects 
are rather small and we have not implemented these effects in our analysis.

We plot in Fig.~1 the resulting 95\% confidence-level exclusion limit on 
$M_X/g_X$ as a function of $\phi$ where $\tan \phi = n_4/n_1$ and the 
normalization $n_4^2 + n_1^2 = 1$ is assumed. This means that instead of 
using the coulplings $g_X n_1$ and $g_X n_4$, we use $g_X \cos \phi$ and 
$g_X \sin \phi$.

\vskip 1.1cm
\input epsf
\begin{figure}[htb]
\begin{center}
\epsfxsize=12cm
\leavevmode
\epsfbox{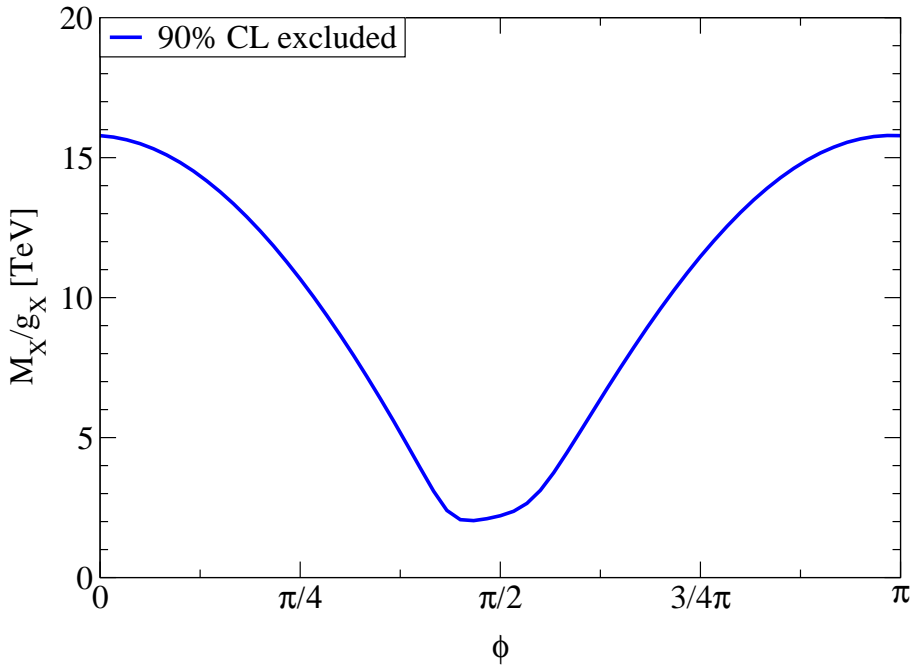}
\end{center}
\vspace*{-5mm}
\caption[] {Lower bound on $M_X/g_X$ versus $\phi = \tan^{-1} (n_4/n_1)$.}
\end{figure}

\noindent \underline{\it Decays of $X$} : 
If the $X$ gauge boson is observed at 
the LHC, then $r = n_4/n_1$ may be determined empirically from its decay 
branching fractions into $q \bar{q}$, $l \bar{l}$, and $\nu \bar{\nu}$, 
which will be proportional to $3(41 - 18 r + 9 r^2)/8$, 
$(81 - 90 r + 41 r^2)/16$, and $r^2$ respectively.  The ratios
\begin{equation}
{\Gamma(X \to t \bar{t}) \over \Gamma(X \to \mu \bar{\mu})} = 
{3(65 - 42 r + 9 r^2) \over 81 - 90 r + 41 r^2}
~~~{\rm and}~~~
{\Gamma(X \to b \bar{b}) \over \Gamma(X \to \mu \bar{\mu})} = 
{3(17 + 6 r + 9 r^2) \over 81 - 90 r + 41 r^2}
\end{equation}
are especially good discriminators \cite{gm08}, as shown in Fig.~2.

\input epsf
\begin{figure}[htb]
\begin{center}
\epsfxsize=12cm
\leavevmode
\epsfbox{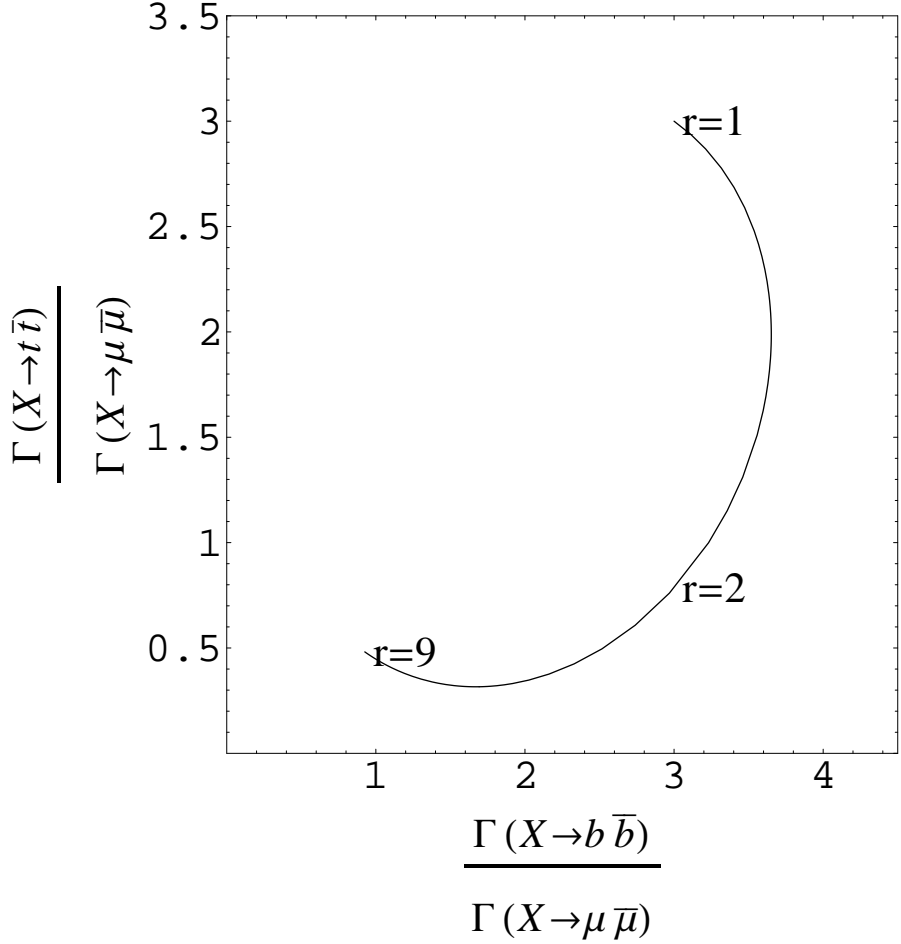}
\end{center}
\vspace*{-5mm}
\caption[] {Plot of $\Gamma(X \to t \bar{t})/\Gamma(X \to \mu \bar{\mu})$ 
versus $\Gamma(X \to b \bar{b})/\Gamma(X \to \mu \bar{\mu})$ as a function of 
$r=n_4/n_1$.}
\end{figure}

\begin{table}[htb]
\caption{$U(1)_X$ content of new particles in Model (C).}
\begin{center}
\begin{tabular}{|c|c|c|}
\hline 
Particle & $U(1)_X$ & $Z_2$ \\
\hline
$N_R, (\Sigma^+,\Sigma^0,\Sigma^-)_{1R,2R}$ & $(3/8)(3n_1+n_4)$ & -- \\ 
$S_{1R}$ & $(1/4)(3n_1+n_4)$ & + \\ 
$S_{2R}$ & $-(5/8)(3n_1+n_4)$ & -- \\ 
\hline
$(\phi^+,\phi^0)_1$ & $(3/4)(n_1-n_4)$ & + \\ 
$(\phi^+,\phi^0)_2$ & $(1/4)(9n_1-n_4)$ & + \\ 
$(\phi^+,\phi^0)_3$ & $(1/8)(9n_1-5n_4)$ & -- \\ 
\hline
$\chi_1$ & $-(1/2)(3n_1+n_4)$ & + \\ 
$\chi_2$ & $-(1/4)(3n_1+n_4)$ & + \\ 
$\chi_3$ & $-(3/8)(3n_1+n_4)$ & -- \\ 
$\chi_4$ & $-(3/4)(3n_1+n_4)$ & + \\ 
\hline
\end{tabular}
\end{center}
\end{table}

\noindent \underline{\it Model with two triplets} : 
We now examine the structure of Model (C) as shown in Table 4.  The fermion 
content is dictated by the anomaly-free conditions for $U(1)_X$ to consist 
of two triplets $\Sigma_{1R,2R}$ and three singlets $N_R, S_{1R,2R}$.  Quarks 
couple to $\Phi_1$ and charged leptons to $\Phi_2$.  However, $(\nu,e)_L$ 
is connected to $N_R$ and $\Sigma_{R}$ through $\Phi_3$, and to $S_{1R}$ 
through $\Phi_1$.  To allow all particles to acquire mass, we add the four 
scalar singlets as shown.  We then have the allowed Yukawa terms $N_R N_R 
\chi_4$, $\Sigma_R \Sigma_R \chi_4$, $S_{1R} S_{1R} \chi_1$, $N_R S_{2R} 
\chi_2^\dagger$, $S_{1R} S_{2R} \chi_3^\dagger$, and the allowed scalar terms 
$\chi_1 \chi_2 \chi_4^\dagger$, $\chi_2^2 \chi_1^\dagger$, $\chi_3^2 
\chi_4^\dagger$, $\chi_1^\dagger \chi_2^\dagger \chi_3^2$, $\chi_2^3 
\chi_4^\dagger$, $\chi_1 \Phi_1^\dagger \Phi_2$, $\chi_3 \Phi_3^\dagger \Phi_2$, 
$\chi_1 \chi_3^\dagger \Phi_1^\dagger \Phi_3$, $\chi_2^2 \Phi_1^\dagger \Phi_2$, 
$\chi_3^\dagger \chi_4 \Phi_3^\dagger \Phi_2$.  Thus the resulting Lagrangian 
has an automatic $Z_2$ symmetry, which implements exactly the proposal of 
Ref.~\cite{m06} for radiative seesaw neutrino masses and dark matter, 
as shown in Fig.~3.  
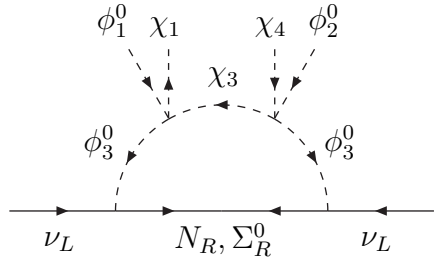
\begin{figure}[htb]
\begin{center}\begin{picture}(500,100)(120,45)
\ArrowLine(270,50)(310,50)
\ArrowLine(390,50)(350,50)
\ArrowLine(310,50)(350,50)
\ArrowLine(430,50)(390,50)
\Text(290,35)[b]{$\nu_L$}
\Text(410,35)[b]{$\nu_L$}
\Text(352,33)[b]{$N_R,\Sigma^0_R$}
\Text(352,97)[b]{$\chi_3$}
\Text(305,70)[b]{$\phi^0_3$}
\Text(396,70)[b]{$\phi^0_3$}
\Text(310,116)[b]{$\phi^0_1$}
\Text(390,116)[b]{$\phi^0_2$}
\DashArrowLine(315,111)(330,85){3}
\DashArrowLine(385,111)(370,85){3}
\DashArrowArc(350,50)(40,120,180){3}
\DashArrowArcn(350,50)(40,60,0){3}
\DashArrowArc(350,50)(40,60,120){3}
\DashArrowLine(370,111)(370,85){3}
\Text(370,116)[b]{$\chi_4$}
\DashArrowLine(330,85)(330,111){3}
\Text(330,116)[b]{$\chi_1$}
\end{picture}
\end{center}
\caption[]{One-loop radiative contribution to neutrino mass.}
\end{figure}
The $3 \times 3$ Majorana neutrino mass matrix receives a tree-level 
contribution from the coupling of $S_{1R}$ to a linear combination of 
$\nu_i$ through $\phi^0_1$, as well as radiative contributions 
from $N_R$ and $\Sigma^0_R$. This is a natural hierachical scenario 
where $\nu_3 = (\nu_\tau - \nu_\mu)/\sqrt{2}$ for example is heavier than 
$\nu_{1,2}$ because the former is the one with a tree-level mass. 

The lightest particle of odd $Z_2$ \cite{dm78} is now a dark-matter candidate. 
However, it is unlikely to be a fermion because it will have $U(1)_X$ gauge 
interactions with nuclei and a cross section proportional to $(g_X/m_X)^4$ 
which is likely to be too big to satisfy the upper limits from direct-search 
experiments and the requirement of the proper dark-matter relic abundance 
through its annihilation.  If it is a scalar boson, such as the lighter of 
Re($\phi_3^0$) and Im($\phi_3^0$) \cite{m06,bhr06,lnot07,glbe07,cmr07} with 
a mass difference greater than about 1 MeV, then it is an acceptable 
candidate because the lighter one is prevented from scattering to the heavier 
one through the $X$ boson kinematically.  On the other hand, the generic 
quartic scalar term for this splitting, i.e. 
$(\lambda_5/2)(\Phi^\dagger \eta)^2 + H.c.$ where 
$\Phi$ is even and $\eta$ odd under $Z_2$, is not available here because of 
the $U(1)_X$ charges.  Nevertheless, splitting does occur in the $4 \times 4$ 
mass-squared matrix spanning Re($\phi_3^0$), Im($\phi_3^0$), Re($\chi_3$), 
and Im($\chi_3$), which is of the form
\begin{equation}
{\cal M}^2 = \pmatrix{m_\phi^2 & 0 & \Delta_2+\Delta_3 & 0 \cr 
0 & m_\phi^2 & 0 & \Delta_2 - \Delta_3 \cr 
\Delta_2 + \Delta_3 & 0 & m_\chi^2 + \Delta_1 & 0 \cr 
0 & \Delta_2 - \Delta_3 & 0 & m_\chi^2 - \Delta_1}.
\end{equation}
Hence $m^2$[Re($\chi_3$)] -- $m^2$[Im($\chi_3$)] = $2 \Delta_1$, and 
$m^2$[Re($\phi_3^0$)] -- $m^2$[Im($\phi_3^0$)] 
= $-4 \Delta_2 \Delta_3/m_\chi^2$.  As for the corresponding relic abundance, 
there will be contributions from the $U(1)_X$ gauge interactions and the 
various allowed Yukawa terms.  Note also that the $Z_2$ symmetry for dark 
matter here is the conserved remnant \cite{ks06,s08,hln08,l08,m08-1,m08-2} 
of $U(1)_X$.

\noindent \underline{\it Conclusion} :  
In this paper, we have discussed some consequences of having 
one or more Majorana fermion triplets $(\Sigma^+,\Sigma^0,\Sigma^-)$ as 
seesaw anchors of neutrino masses in the context of an U(1) extension of 
the SM. The associated neutral gauge boson $X$ has prescribed couplings to 
the usual quarks and leptons in terms of $g_X$ and $\phi = \tan^{-1} 
(n_4/n_1)$.  The exclusion limit on $M_X/g_X$ from low-energy data has 
been obtained, showing that $X$ may be accessible at the LHC if $g_X$ is 
of order $g_Z$.  In the case of one triplet, i.e. Model (B), one Higgs 
doublet couples to quarks and the other to leptons. In the case of two 
triplets, i.e. Model (C), there is a third scalar doublet, which allows 
for the natural implementation of radiative neutrino masses and dark matter.

\noindent \underline{\it Aknowledgements} : 
This work was supported in part by the U.~S.~Department of Energy under
Grant No. DE-FG03-94ER40837, by UNAM as DGAPA--PAPIIT project IN115207, 
and by CONACyT (M{\'e}xico) as project 82291--F.  We thank the Institute 
for Nuclear Theory at the University of Washington for its hospitality 
and the Department of Energy for partial support during the completion 
of this work.

\bibliographystyle{unsrt}

\end{document}